\documentstyle[12pt,epsfig,aaspp4]{article}

\def\deg{\ifmmode^{\circ}\else$^{\circ}$\fi} % overwrites \deg in LaTeX
\def\min{\ifmmode^{\prime}\else$^{\prime}$\fi}
\def\sec{\ifmmode^{\prime\prime}\else$^{\prime\prime}$\fi}
\def\fdeg{\ifmmode.\!\!^{\circ}\else$.\!\!^{\circ}$\fi}
\def\fmin{\ifmmode.\!\!^{\prime}\else$.\!\!^{\prime}$\fi}
\def\fsec{\ifmmode.\!\!^{\prime\prime}\else$.\!\!^{\prime\prime}$\fi}
\def\th{\hbox{$^{\rm h}\,$}}
\def\tm{\hbox{$^{\rm m}\,$}}

\def\fs{\hbox{$.\!\!^{\rm s}$}}

\def\etal{{\it et~al.\ }}

\begin{document}

\title{THE RESOLVED OUTER POPULATION OF NGC6822 WITH WFPC2}

\author{J. B. Hutchings\altaffilmark{1}} \affil{Dominion Astrophysical
Observatory,
National Research Council of Canada,\\ Victoria, B.C. V8X 4M6, Canada}
\author{B. Cavanagh}
\affil{Department of Physics and Astronomy, University of Victoria, 
B.C. V8W 3P6, Canada}
\author{L. Bianchi\altaffilmark{1}} \affil{Center for Astrophysical Sciences,
Dept of Physics and Astronomy, Johns Hopkins University, Baltimore, MD 21218}

\altaffiltext{1}{Observer with the NASA/ESA {\it Hubble Space
Telescope} at the Space Telescope Science Institute, which is 
operated by the Association of Universities for Research in Astronomy, Inc.,
under NASA contract NAS 5-26555.}

\begin{abstract}
We present F336W ($U$), F439W ($B$), F555W ($V$), and F675W ($R$) Wide Field
Planetary Camera 2 (WFPC2) photometry of two outer regions of the Local 
Group dwarf irregular galaxy NGC6822. The NE region is $\sim$13 arcmin from
the galaxy centre, while the W region lies 10 arcmin out, and within the 
wispy low surface brightness outer regions of the galaxy. The fields are
not crowded and contain few NGC 6822 stars. We discuss errors and
uncertainties and find that
the W region contains a main sequence that extends to stars
of about 2 M$_{\odot}$, with an age of about 200 Myr. The NE region
has no main sequence or stars younger than 1 Gyr, but does contain some 
luminous red stars that are not matched in the W field. These stars
are not clumped in the field. The results suggest that the W region
may be a trace of a tidal event that triggered the current star-formation
in this isolated galaxy.

\end{abstract}

\keywords{Galaxies: stellar content, galaxies: local group}

\section{INTRODUCTION}
The Local Group dwarf irregular galaxy NGC6822 was the first extra-Galactic
object to have its distance measured by Hubble (1925) using the period-luminosity relation of Cepheids.  Since these original observations,
NGC6822 has been studied in varying
degrees of detail. This galaxy has been classified as a type Ir IV-V by 
van den Bergh (1968), which is a similar classification as the Small Magellanic
Cloud.  The distance modulus has been more accurately determined since Hubble's 
initial calculation of 21.65 (corresponding to a distance of 214 kpc); current
values are close to 23.5, which is a distance of 500 kpc (McAlary et al 1983).
NGC6822 has optical dimensions of 6\min $\times$ 11\min, which corresponds to 
a size of 1.7 $\times$ 3.2 kpc at a distance of 500 kpc. Galaxy parameters
are given in Table 1. Roberts (1972) showed that the
optical core is embedded in a large gaseous envelope with dimensions 42\min 
$\times$ 89\min (12.2 $\times$ 25.9 kpc at 500 kpc). Inspection of the
DSS image of the galaxy shows that the W side has extensive faint streaky
light that is not present on the E side. It is of interest to know whether
this arises from stars belonging to the galaxy. If so, they may be the
remnants of a tidal event that has triggered the current star-formation
activity in this apparently isolated galaxy.

Many of the previous photometric surveys of NGC6822 have studied either specific
objects within the galaxy, or the galaxy as a whole.  NGC6822 has many \ion{H}{2} regions,
most of which lie along the northern edge of the galaxy, and which have been 
studied extensively.  OB associations in the galaxy have also been studied
(Massey et al 1995).  
Stars found in these regions are mostly young, high-luminosity stars.  Very
few studies have examined the outer regions of NGC6822 to probe its 
stellar population. For example,  Massey's (1998) 
control field is closer to the galaxy than the fields of this study.  
A `halo' population was found in the Local Group 
Im IV-V dwarf irregular galaxy WLM by Minniti and Zijlstra (1996), so a 
similar population in NGC6822 seems likely.  

\section{OBSERVATIONS}

Observations were obtained with the Wide Field Planetary Camera 2 (WFPC2)
on the {\it Hubble Space Telescope} ($HST$) as coordinated parallels
with spectroscopy of OB stars in selected associations (program GO6567). 
The spacecraft roll angles were specified to maximise overlap in two areas 
to the E and W of the main galaxy body, using five different 
filters. For each pointing we obtained exposures with two different filters, 
in order to obtain the maximum filter coverage for the overlapping areas, 
and providing one color for each complete WFPC2 field.
Details are shown in Table 2. 

Data sets 1 \& 2 (the `East' region)
overlap some 13 arcsec NE of the galaxy centre, and data sets 3, 4, \& 5 
(the `West' region) overlap some 10 arcsec W of the galaxy centre. This
enables photometry in four wavelength bands for a subset of resolved stars.
Figures 1 \& 2 show the positions of the fields on the sky.

The data were reduced using profile-fitting photometry, and examined for
individual pointings, and then for the overlap regions of nearby pointings.
The W sets (3, 4, 5) lie within the streaky extended flux of the galaxy
while the E region (1, 2) appear to lie clear of any galaxy light seen in the DSS. Aperture corrections were made to determine correct 
zeropoints and offsets, followed by foreground star decontamination and
de-reddening corrections, as described in the sections following.

\section{PHOTOMETRY}

\subsection{Individual Sets}

All individual sets of WFPC2 images were reduced using
DAOPHOT (Stetson 1987) under IRAF\footnote{IRAF is distributed by the National 
Optical Astronomy Observatories, which are operated by the Association of
Universities for Research in Astronomy, Inc., under cooperative agreement
with the National Science Foundation.} using the following technique:
 
(1) the FIND
routine was used to find starlike objects exceeding the local background
by four standard deviations in the images observed at the higher wavelength 
(i.e., for data set 1, the observation with the F555W filter was used); 

(2) the PHOT 
routine was used to determine initial estimates of the magnitude of each 
starlike object by aperture photometry;

 (3) a point-spread function was determined by manually
selecting stars in each frame, then using a Moffat function with $\beta$=1.5
to fit to the stellar images to a radius of 10 pixels; 

(4) the ALLSTAR routine was used to 
obtain profile-fitting magnitudes for the stars in the frame;

 (5) aperture
photometry of the images at lower wavelength was performed using the
coordinate list obtained from the higher wavelength observation, using the
PHOT routine; and 

(6) the ALLSTAR routine was used to obtain profile-fitting
magnitudes for stars in the frame observed at the lower wavelength using the
PSF obtained for the higher wavelength observation. 

  The sky levels were determined from simple statistics on the entire
images, and checked against the final star-subtracted images. The fields
are not crowded so that the sky level was always well determined and not 
a source of photometric error. Once magnitudes for each
wavelength were obtained, the resulting tables were merged according to star
ID number; if a starlike object was detected and measured in both wavelengths,
then there is a good chance that this object is a star. Chance alignments with
cosmic rays were eliminated by rejecting objects with a {\it CHI} fitting
parameter greater
than 1.5 in the lower wavelength images.  A random inspection of objects thus
rejected showed that they were indeed cosmic rays and not stars.  
This threshold also rejects stars that have saturated the WFPC2 detectors
and would have otherwise been rejected. 

ALLSTAR estimates measuring errors for all objects. These reach a mean of 
0.2 mag at the following magnitudes (and filters): 24.7 (F675W), 
24.5 (F555W), 23.2 (F439W), 22.3 (F336W). Objects with errors
in the higher wavelength band greater than 0.25 mag were also rejected.

\subsection{Overlapping Sets}

The overlapping sets were reduced by the method described above, with some
modifications.  First, images were aligned using reference stars in each
overlapping image.  To ensure the photometry does not change when images
are shifted and rotated, aperture photometry was done on the reference stars
in the pre- and 
post-alignment images; results agreed to within 0.02 magnitudes, below the
average errors obtained from ALLSTAR.  After the images were aligned, 
magnitudes for all stars were obtained using the above techniques.  Once
magnitudes were
determined, objects appearing at the same location (within a 3-pixel fitting
radius) in all four wavelengths were kept.  Again, there is a chance of 
alignment with cosmic rays, so objects with a {\it CHI} fitting
parameter greater than 1.5 were rejected.  Objects with errors
in the higher wavelength band greater than 0.25 mag were also rejected.

\subsection{Aperture Corrections and Errors}

Because the ALLSTAR profile-fitting routine uses, as a basis for its magnitudes,
a preliminary crude estimate of the magnitude obtained using a small synthetic
diaphragm, chosen to avoid contamination from neighbouring stars, the magnitudes
it obtains will be offset from the true instrumental magnitudes by some amount.
This offset was determined by performing aperture photometry on uncrowded stars,
primarily those stars used to determine the PSF in step (3), above.  For 
observations at lower wavelengths (for which no PSF was determined), uncrowded
stars were chosen by hand.  In addition, because the zeropoint of each WFPC
chip as given in the WFPC2 Instrument Handbook (Burrows 1995) varies slightly,
photometric offsets were determined for each chip individually.  The final
magnitudes were calculated from these offsets.

Fields 3 and 5 have overlapping F555W observations and fields 4 and 5 have
overlapping F336W observations. This enabled us to make empirical checks on 
the agreement between measures of the same stars in individual and combined
images. The differences in stars of magnitudes to 25 are less than the errors 
given by ALLSTAR. Thus, we adopt the errors given by the program, which 
are very similar for individual and combined frames with these exposures
and adopted limiting signals. 

We made no correction for telescope breathing, although the exposures were
short compared with the telescope breathing cycle time. However,
the PSF radius and co-addition position limits were large enough that 
this would be an effect of a few percent
or less, and thus not a significant source of photometric error.

Figures 3, 4, and 5 display selected aperture-corrected colour-magnitude
diagrams. The magnitudes obtained after this step will be referred to as 
the {\it observed magnitudes}. Figure 5 shows error bars for all stars, 
which are the same as in Figure 3. Figure 4 shows errors bars typical of 
3 places in the diagram, as placing errors bars on all stars makes the 
diagram too crowded to see easily.

\section{FOREGROUND CONTAMINATION REMOVAL}

Because NGC6822 lies at a low galactic latitude ($b_{1950} = -18\fdeg40$),
and because our fields lie outside the dense central regions of NGC 6822,
contamination from foreground Galactic stars is a concern. Wilson (1992) 
used control fields that were located 30\min east and west of NGC6822; this 
distance would be ideal for this study.  A search for WFPC2 images within 
one degree of NGC6822 was unfruitful; the only objects observed in this
region are specific objects within NGC6822 (e.g., OB associations, \ion{H}{2}
regions). Thus, for decontamination, archival WFPC2 images of the 
quasar Q1240+1754, taken with the
F439W and F675W filters were measured. While they are not ideally placed
to match the immediate foreground environment of NGC 6822, they are the
only images with some of the same filters and exposures (350 - 400 sec),
with suitable CR-split readouts. 

The two main fields lie in regions of different light level associated with 
NGC 6822, and their C-M diagrams also differ in the sense of indicating
that the W field contains more NGC 6822 stars. Thus, we also used the 
NE field as the `foreground' to decontaminate the W field. 

To remove foreground stars from the colour-magnitude diagrams of NGC6822,
we first established that the limiting magnitude and errors from the
photometry were the same, and also compared the total numbers of stars
measured in the two fields. The QSO field was less heavily populated,
being at higher galactic latitude ($b \approx 80\deg$), so the density
of foreground stars was lower. This was somewhat offset in the overlap fields
by their smaller area. In each field being corrected, total numbers 
of stars were scaled (always upwards) in the control field to match those 
in the field of
interest, at the bright end of the distribution. These comparisons were
made with NGC 6822 fields in the NE and W, and were consistent. The scaled
QSO field file was increased by adding the same extra fraction of stars
in half-magnitude bins, and giving the extra stars magnitudes that differ
by 0.01 mag. A script was then
written to perform the following: a star in the foreground field is chosen.
If there is a star in the background field (the field of interest) such 
that if

\begin{equation}
\sqrt{\Delta_{color}^2 + \Delta_{magnitude}^2} < x,
\end{equation}

\noindent (where $\Delta_{color}$ is the difference in colour between the foreground 
and background stars, $\Delta_{magnitude}$ is the difference in magnitude
between the foreground and background star, and $x$ is an arbitrary 
fitting radius)
then the foreground star cancels out the background star, and both
stars are removed, before proceeding to the next comparison.  

Because the QSO control field was only observed with filters
F439W and F675W, only these colours were used in this exercise.
When using the NE fields as control for the W fields, only the relative
areas of sky were used in the number scaling.

The results of the foreground decontamination, using a fitting radius 
$x=0.1$ mag, are shown in Figures 3 and 5. Our approach to foreground
removal is simple, but all that is warranted by the small numbers of stars
and photometric errors we have. However, the principal result is the
distribution of bright stars in the decontaminated diagrams, and these
agree well between the two methods (scaled QSO field and use of NE field),
as well as being visible in the uncorrected diagram (Fig 4). Thus, the
decontamination process is not critical to the main result we discuss 
in the next section.   

\subsection{De-reddening Corrections}

The reddening towards NGC6822 has been determined many times in the past,
with varying results.  Kayser (1967) found, by using two-colour photographic
photometry of field stars, a mean foreground reddening of E($B-V$) =
0.27$\pm$0.03. Hodge (1977) determined that the average value for the 
reddening towards NGC6822 was E($B-V$) = 0.28$\pm$0.03.  van den Bergh 
\& Humphreys (1979) found E($B-V$) $\sim$ 0.3 for field stars around the galaxy.
Massey \etal (1995) showed that there is a spatial dependence of the
reddening; reddening is smallest [E($B-V$) $\approx$ 0.26], on the east and
west extremes of the galaxy, and as high as E($B-V$) $\approx$ 0.45 near 
the centre. Because the two regions examined in this paper lie well outside 
the body of the galaxy, we have used a constant reddening value of 
E($B-V$)=0.26 which we regard as almost all foreground. In addition, the
ratio E($U-B$)/E($B-V$)$\approx$0.72, so a reddening value of E($U-B$)=0.19 is
used.  A distance modulus of 23.49 was used to determine absolute
magnitudes.

The dereddening was done after the foreground decontamination,
since the reddening of foreground stars will be close to zero in both
the NGC 6822 and control fields.

\section{RESULTS AND DISCUSSION}

   In discussing the resulting diagrams we have overplotted the
isochrones of Bertelli et al (1994). We find that the isochrones for
solar abundance fit the data better than the LMC abundance, although
some intermediate value is indicated. We discuss this further below.
Figure 5 shows the comparison
of models, and for simplicity, only solar abundance models are plotted
in Figures 3 and 4. We have converted the model colours from UBVR to the
WFPC2 filter passbands, following the WFPC2 handbook. 
The data are dereddened as above before plotting.

  Figures 3, 4, and 5 use the largest dataset that contains the filters
illustrated. These do not differ detectably from subsets that overlap in 
other filters. Because of the large number of images and datasets with
overlap, we do not present tables of individual measures. However,
the first author will provide tables to those interested.

Comparisons to similar colour-magnitude diagrams for other dwarf irregular galaxies such as Pegasus 
(Gallagher \etal 1998) and Sextans A (Dohm-Palmer \etal 1997), and with
previous studies of NGC6822 (see e.g., Wilson 1992, Massey \etal 1995, \& Gallart \etal
1996) show two major differences, the most apparent being a relative lack of
a main-sequence branch in these new results.  This is because 
the previous studies are in the body of the galaxy, and some 
targeted specific star-formation regions (OB associations primarily).
Gallart \etal (1996) showed that the [($B-V$), $V$] colour-magnitude diagram 
is useful for
studying the youngest stars (ages $<$ 250 Myrs), whereas the [($V-I$), $I$]
colour-magnitude diagram contains stars of all ages, up to $\approx$ 15 Gyrs.

    In the outer fields we have observed, there are very few stars, but
their age is of interest in establishing their origin and connection with
current star-formation. While the numbers are very small, the stars of
particular interest are well detected and measured, and thus we feel
their implications are worth discussing.

   Figure 3 shows the result of using the E field as control for the W
field (open circles). This should reveal the difference in populations
in the two outer fields, and remove foreground objects. We choose
to decontaminate this way as the E field clearly has stars that may be 
NGC 6822 main sequence which are not present in the W field. As noted,
the W field is also closer and within the irregular outer faint flux 
associated with the galaxy. However, the filled 
circles are stars in the E field that have no counterpart in the W field.

   Figure 4 shows the E and W fields without foreground correction in 
the two filters with our longest wavelength separation. The same differences
in population are seen. 

   Figure 5 shows both E and W fields combined, corrected by the QSO 
control field with these two filters. 
 
   In all these diagrams, the W field contains a number of brighter blue 
stars not seen (at all) in the E field, indicating a NGC 6822 population 
of age some 100 Myr. For Poisson statistics, the excess of main sequence
objects in the W field (and the excess of bright red stars in the E
field) is signficant at  $>3\sigma$ level. The redder population
of bright stars in the E field is harder to explain.  They are spread over
the field, so are not associated with a cluster within NGC 6822. The most
luminous stars could be massive NGC 6822 stars in post red giant evolution,
whose duration (and hence observed population) is comparable to the main
sequence phase. However, the numbers would imply a top-heavy IMF (and 
small age spread of $<$ 30 Myr), since there are no main sequence stars 
in this field. Also, there are too few red giant stars (particularly
as some might be old metal-poor red giant stars reaching M$_V\sim$-2.5 
and B-V $\sim$1.6), and
too many in the region between the main sequence and the red giants at
ages around 60Myr. These stars may also 
be the giant branch population of stars in our galaxy halo that are unevenly
distributed between our E and W fields.

    We note that the solar abundance models fit the reddest colours
better than the LMC abundance models (see figure 5). In addition, the 
colours of the bright intermediate stars lie closer to the solar blue
loop limit than for LMC abundance. Although, as noted in the previous 
paragraph, we are not clear what the connection is, overall, the models 
appear to fit best for an abundance closer to solar than LMC. While
this is not a strong result in view of the small numbers of stars
and the outer regions sampled, it differs from the LMC type
abundance found in the main galaxy. Thus, the explanation may lie
more in understanding the evolutionary status of the red stars than in
their abundances.

The bulk of our main sequence NGC 6822 stars, in the W region, are 
comparable in age 
with the the oldest stars seen  by Gallart \etal (1996) in their 
[($B-V)_0, M_V$] colour-magnitude diagram, at about 200 Myrs.  To fully
examine the older population of the halo of NGC6822, further observations in 
redder filters (specifically the F814W ($I$) filter) would be needed.  The 
so-called {\it blue plume} at ($m_{439} - m_{555})_0 = 0$ is composed of
both main sequence and blue-loop stars.  Gallart \etal (1996) were able
to differentiate between these two components by noticing a change in 
stellar density, but because of the low numbers of stars populating the
blue plume in this study no such differentiation can be made.  
The E region shows no sign of a blue plume at ($m_{439} - m_{555})_0 = 0$.

These results may also be compared with the [($B-V), V$] colour-magnitude
diagram of the Pegasus dwarf irregular galaxy by Gallagher \etal
(1998), specifically that of the WF3 chip.  In the Gallagher \etal study, 
the bulk of
the galaxy lay in the PC1, WF2, and WF4 chips, and the WF3 chip contained
mostly the outlying regions of this galaxy.  The colour-magnitude diagram
for this chip shows a very sparsely populated main sequence branch, very
similar to what is seen in our diagrams.  A main sequence branch is seen in 
the colour-magnitude diagrams for the other three chips, which is expected, 
as these chips cover the body of the galaxy.

We suggest that the blue stars seen in the W field represents a younger
population that is connected with the streaky faint light seen on the W side
of the galaxy, and not seen to the E. Both phenomena may have arisen from a
tidal event that disrupted the older population in this direction and also 
led to the current star-formation activity in the whole galaxy, and also
traced in the debris to the W. Wider field imaging with deeper exposures
would be needed to explore this possibility further.

We acknowledge the critical reading and helpful suggestions of a referee.

\newpage

\centerline{\bf Captions to figures}

\figcaption{WFPC2 fields 1 \& 2 (the `East'
region).  The background image comes from the STScI
Digitized Sky Survey, and measures approximately 14\fmin5 on each side. 
North is up and E to the left.}

\figcaption{As in Fig. 1, but for fields 3, 4, \& 5 (the `West' region).}

\figcaption{NGC 6822 W field after foreground subtraction using E field.
The filled symbols are stars in the E field that were not matched with
any W field stars in the process. Superposed tracks are Bertelli et al
(1994) isochrones for ages shown.}

\figcaption{Widest colour separation plot of stars from overlap regions
in E and W fields. No foreground subtraction has been applied. Note the
presence of some main sequence stars in the W field.}

\figcaption{Combined results from E and W fields with foreground subtraction
using data with these filters in high latitude QSO field. The presence
of NGC 6822 main sequence stars is evident. The solid isochrone
tracks are for solar
abundance models, and the dashed and dotted ones are for LMC abundance.}

\clearpage

\begin{deluxetable}{rlc}
\singlespace
\tablecaption{Parameters for NGC6822.}
\tablehead{ 
 & & \colhead{Ref.} }
\startdata
$\alpha_{2000}$ & $19\th44\tm56\fs14$ & \\
$\delta_{2000}$ & $-14\deg48\min05\fsec5$ & \\
$l_{1950}$ & $25\fdeg34$  \\
$b_{1950}$ & $-18\fdeg39$  \\
$(m-M)_0$ & $23.49\pm0.08$ & 1 \\
E($B-V$) & 0.26 & 2 \\
\enddata
\tablerefs{(1) Gallart \etal (1996); (2) Massey \etal (1995). }
\end{deluxetable}

\begin{deluxetable}{crrrc}
\tablewidth{0pt}
\singlespace
\tablecaption{Journal of observations (GO6567)}
\tablehead{
\colhead{Data set} & \colhead{R.A. (J2000)} & \colhead{Decl. (J2000)} & \colhead{Filter} & \colhead{Exposure time (s)}
}
\startdata
U3860103T (1) & 19 45 36.453 & -14 39 14.784 & F555W & 300 \\
U3860105T (1) & 19 45 36.453 & -14 39 14.784 & F439W & 350 \\
U3860203M (2) & 19 45 31.972 & -14 38 15.956 & F675W & 300 \\
U3860205M (2) & 19 45 31.972 & -14 38 15.956 & F336W & 350 \\
U3860303T (3) & 19 44 18.284 & -14 45 19.498 & F555W & 300 \\
U3860305T (3) & 19 44 18.284 & -14 45 19.498 & F439W & 350 \\
U3860403T (4) & 19 44 16.404 & -14 45 32.303 & F675W & 300 \\
U3860405T (4) & 19 44 16.404 & -14 45 32.303 & F336W & 350 \\
U3860503T (5) & 19 44 15.706 & -14 46 12.936 & F555W & 300 \\
U3860505T (5) & 19 44 15.706 & -14 46 12.936 & F336W & 350 \\
\enddata
\end{deluxetable}


\newpage

\begin{references}

\reference{} Bertelli, G., Bressan, A., Chiosi, C., Fagotto, F., \& Nasi, E.,
1994, AASuppl, 106, 275

\reference{} Burrows, C.J. (Ed.) 1995, Wide Field and Planetary Camera 2
Instrument Handbook (STScI)

\reference{} Dohm-Palmer, R.C., Skillman, E.D., Saha, A., Tolstoy, E.,
Mateo, M., Gallagher, J., Hoessel, J., Chiosi, C., \& Dufour, R.J. 1997 AJ, 114,
2514

\reference{} Gallagher, J.S., Tolstoy, E., Dohm-Palmer, R., Skillman, E.D.,
Cole, A.A., Hoessel, J.G., Saha, A., \& Mateo, M. 1998, AJ, 115, 1869

\reference{} Gallart, C., Aparicio, A., \& V\'{i}lchez, J.M. 1996, AJ, 112, 1928

\reference{} Hodge, P.W. 1977, ApJS, 33, 69

\reference{} Hubble, E.P. 1925, ApJ, 62, 409

\reference{} Kayser, S.E. 1967, AJ, 72, 134

\reference{} Massey, P. 1998, ApJ, 501, 153

\reference{} Massey, P., Armandroff, T.E., Pyke, R., Patel, K., \& Wilson, C.D.
1995, AJ, 110, 2715

\reference{} McAlary, C.W., Madore, B.F., McGonegal, R., McLaren, R.A.,
\& Welch, D.L., 1983, ApJ, 273, 543

\reference{} Minniti, D., \& Zijlstra, A.A. 1996, ApJL, 467, 13

\reference{} Roberts, M.S. 1972, IAU Symposium 44, D.E. Evans (Ed.)

\reference{} Stetson, P.B. 1987, PASP, 99, 191

\reference{} van den Bergh, S., 1968, Communications of the David Dunlop
Observatory, No. 195

\reference{} van den Bergh, S.\& Humphreys, R.M. 1979, AJ, 84, 604

\reference{} Vaucouleurs, G. de, Vaucouleurs, A. de, \& Corwin, H.G. 1976, 
Second Reference Catalogue of Bright Galaxies (University of Texas Press,
Austin)

\reference{} Wilson, C.D. 1992, AJ, 104, 1374

\end{references}
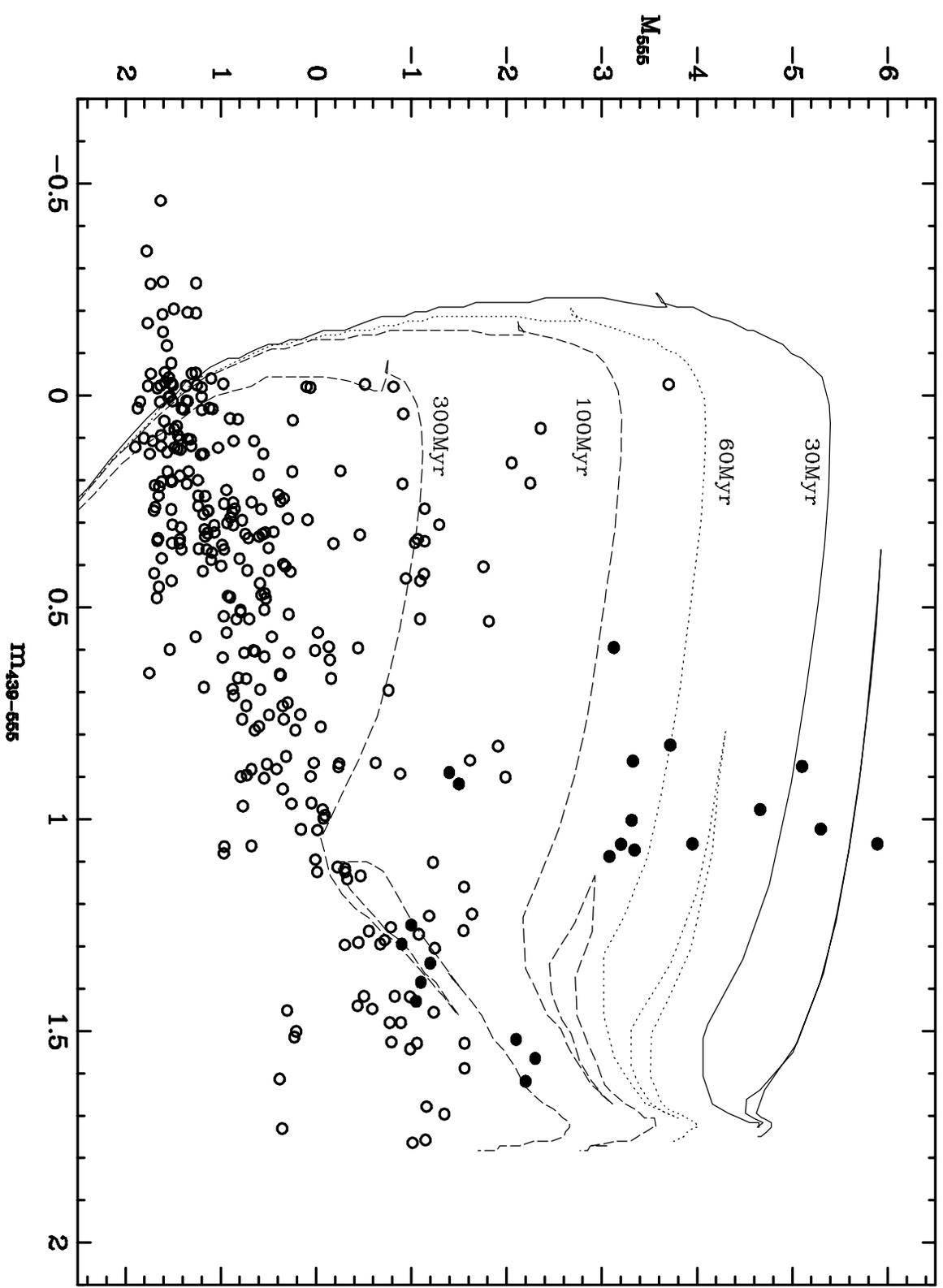
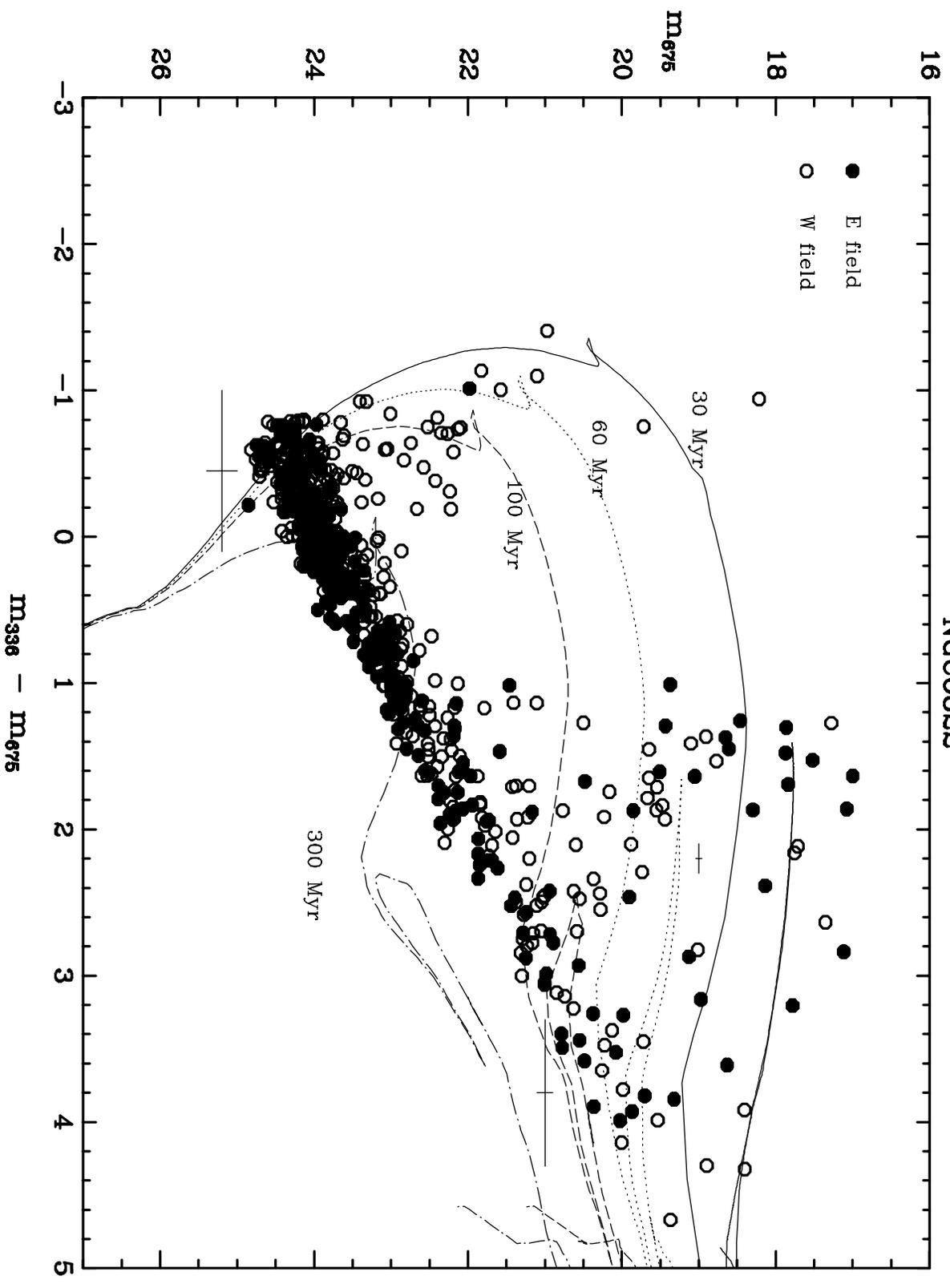
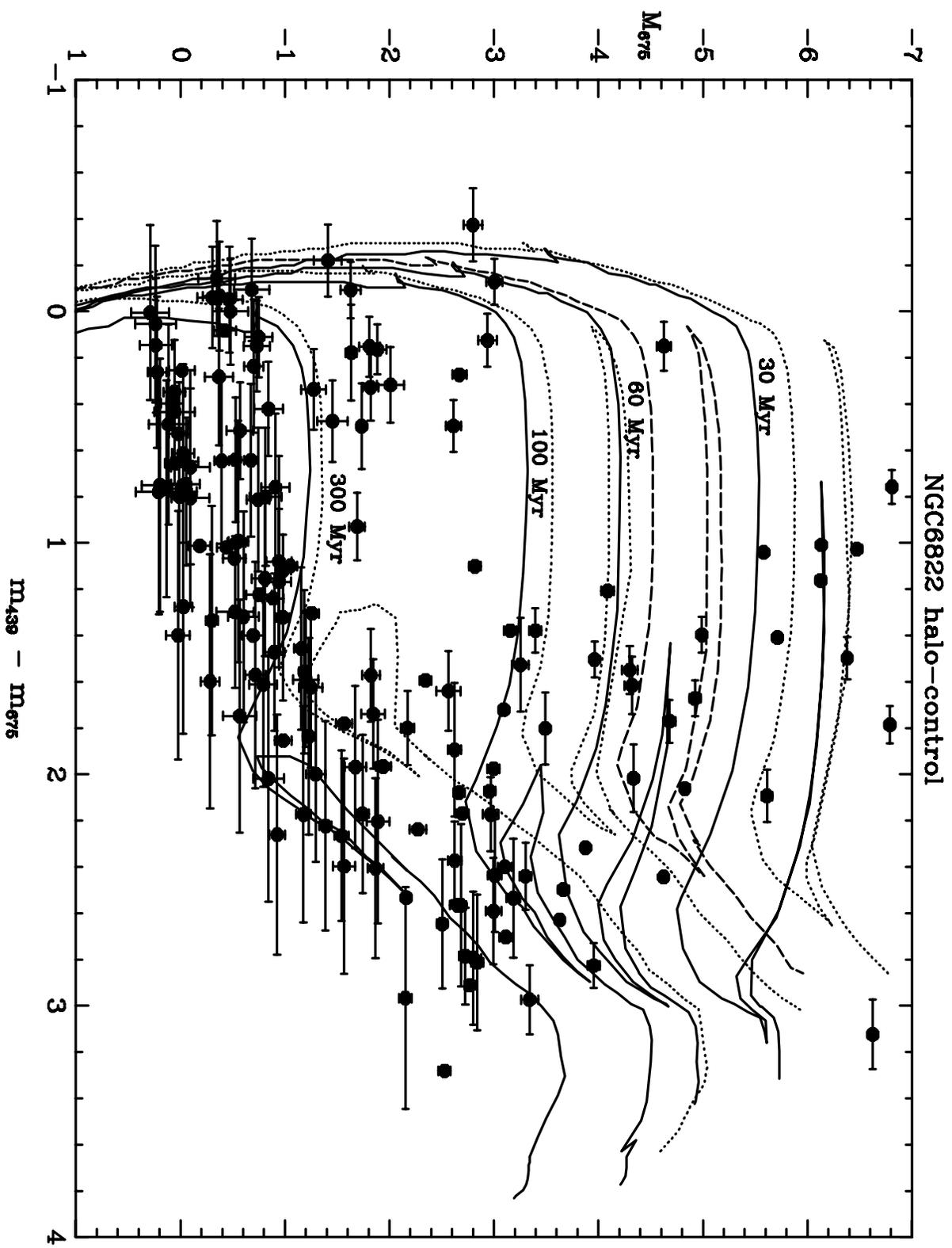
\end{document}